\documentclass[useAMS]{mn2e}

\usepackage{graphicx}
\usepackage{amssymb}

\font\tenbg=cmmib10 at 10pt

\def \rvecphi{{\hbox{\tenbg\char'036}}}

\title{Rossby Wave Instability with Self-Gravity}
\author[R.V.E. Lovelace \& R.G. Hohlfeld]{R.V.E. Lovelace$^{1}$ \& R.G.  Hohlfeld$^{2,3}$\\
$^{1}$Department of Astronomy, Cornell University, Ithaca, NY
14853; email: lovelace@astro.cornell.edu \\
$^{2}$Center for Computational Science,  Boston University,
3 Cummington Street, Boston MA 02215; \\
$^3$Wavelet Technologies, Inc.
664 Pike Ave., Attleboro MA  02703; RHohlfeld@wavelettech.com\\}

\begin{document}

\maketitle

\begin{abstract}

   The  Rossby wave instability (RWI) in non-self-gravitating discs can be
triggered by a bump at a radius $r_0$ in the disc surface
mass-density (which is proportional to the inverse potential vorticity).
    It gives rise to a growing non-axisymmetric perturbation
[$\propto \exp(im\phi)$, $m=1,2..$]  in the vicinity
of $r_0$ consisting of {\it anticyclonic} vortices 
which may facilitate  planetesimal growth in protoplanetary discs.
    Here, we analyze a continuum of thin disc models ranging 
from self-gravitating to non-selfgravitating.  
     The key quantities
determining the stability/instability are: (1) the parameters of the bump
(or depression) in the disc surface density, (2) the Toomre
$Q$ parameter of the disc (a non-self-gravitating disc
has $Q\gg1$), and (3) the dimensionless azimuthal wavenumber of the 
perturbation $\overline{k}_\phi =mQh/r_0$, where $h$ is the half-thickness of the disc.   
      For discs stable to axisymmetric perturbations ($Q>1$),  the self-gravity 
has a significant role   for $\overline{k}_\phi  <  \pi/2$ or $m<(\pi/2) (r_0/h)Q^{-1}$;   instability may occur  for a depression or groove in the surface density if $Q\lesssim 2$.
  For  $\overline{k}_\phi  >  \pi/2$ the self-gravity is not important, and
instability may occur at a bump in the surface density.
Thus, for all mode numbers $m \ge 1$, the self-gravity is unimportant
for $Q > (\pi/2)( r_0/h)$.   We suggest that  the self-gravity be included
in simulations for cases where $Q< (r_0/h)$.

\end{abstract}


\section{Introduction}
   
    The  theory of the  Rossby wave  instability (RWI) 
was developed by Lovelace et al. (1999) 
 and  Li et al. (2000) for thin accretion discs with negligible self-gravity  and earlier by Lovelace \& Hohlfeld (1978) for thin disc galaxies where the self-gravity may or may not be important.
      In the first case the instability can occur if there
is a bump as a function of radius in the inverse potential vorticity $F(r)\propto
\Sigma \Omega [({\bf \nabla \times u})\cdot\hat{\bf z}]^{-1}$
at some radius $r_0$, where $\Sigma$ is the surface mass density,
$\Omega $ the angular velocity, and ${\bf u}$ the flow velocity of the disc.
    Such a  bump could arise at the radial boundary of the ``dead zone''
(Varni\`ere \& Tagger 2006;  Lyra et al.  2009; Crespe et 
al. 2011 ).  
   This zone can arise from the suppression of 
the  magnetorotational instability  by low ionization
 inside the disc (Gammie 1996).
    In turn, the bump in $F(r)$  can give rise to the
exponential growth of non-axisymmetric perturbations in the vicinity
of $r_0$ consisting of {\it anticyclonic} vortices.
    Such vortices can act to concentrate dust grains in
their centres and thereby accelerate the formation of macroscopic
planetesimals (Barge \& Sommeria 1995; Tanga et al. 1996; 
Bracco et al. 1999; Godon \& Livio 2000;
Johansen et al. 2004; Heng \& Kenyon 2010).
   Two-dimensional hydrodynamic simulations of the RWI
instability in discs were done by Li et al. (2001) and recently
in three dimensions  (e.g.,  Meheut et al. 2010, 2012a).  
    The theory
of the instability in three dimensions has been studied by
Meheut et al. (2012b) and Lin (2012). 
   The instability is also predicted to occur in strongly non-Keplerian discs 
with regions where $d\Omega/dr >0$  which exist around rotating magnetized stars (Lovelace et al. 2009).   
     In the RWI the wave is radially trapped within the disc which
is different from the Papaloizou and Pringle (1984; 1985) instability
where the wave is trapped between the inner and outer radii of
a disc or torus.
         The RWI has an important role in the accretion-ejection 
instability of discs proposed by Tagger and collaborators
(e.g., Tagger \& Varni\`ere 2006; Tagger \& Melia 2006; Tagger \& Pellat 1999).

    The RWI in disc galaxies consisting mainly of
stars has been studied  using $N-$body simulations
(Sellwood \& Kahn 1991; and Sellwood 2012).  
   When the  self-gravity of the disc is important,
instability  occurs at radii where $F(r)$ has a minimum, that is,
a groove (Lovelace \& Hohlfeld 1978; Sellwood \&Kahn 1991).
     We do not consider the instability of self-gravitating gas discs
for conditions where the gas cooling is important (Gammie 2001).

    Saturation of the exponential growth of the instability is predicted
to occur when the time-scale for a fluid particle to orbit the center of
a vortex  is comparable to the wave's growth time-scale 
(Lovelace et al. 2009).   Simulations by  Meheut et al. 2012c) support
this conclusion.

   Section 2 develops the theory, applies it to 
radially localized perturbations, and considers 
axisymmetric and nonaxisymmetric modes.
Section 3 gives sample results for cases where
self-gravity is important and where it is negligible,
and Sec. 4 gives a numerical example relevant
to forming protostar.
   Conclusions are given in Sec. 4.

\section{Theory}

      We consider the stability of
of a thin self-gravitating disc of
equilibrium surface mass density $\Sigma(r)$
with a $(r,\phi,z)$ coordinate system.
   The equilibrium  has
the flow velocity
${\bf u} = u_\phi(r) \hat{\rvecphi~}=r\Omega(r) \hat{\rvecphi~}$.  
   That is, the accretion velocity $u_r$ and the vertical
velocity $u_z$  are assumed negligible compared with $u_\phi$.
     The equilibrium flow satisfies $-\Sigma r \Omega^2=
-dP/dr-\Sigma \nabla \Phi$, where  $P$ the vertically
integrated pressure and $\Phi$ the gravitational potential.
This potential is give by $\nabla^2 \Phi =4\pi G\Sigma \delta(z)$,
where $G$ is  the gravitational constant.

The perturbed quantities are:
the density,
$\tilde{\Sigma} = \Sigma +
\delta \Sigma(r,\phi,t)$;
the pressure is
$\tilde{P} = P+\delta P(r,\phi,t)$;
the  flow
velocity is $\tilde{\bf u} =
{\bf u} +\delta {\bf u}(r,\phi,t)$ with
${\bf \delta u} =
(\delta u_r,\delta u_\phi,0)$. 
    The equations for the
perturbed flow are
$$
{D \tilde{\Sigma}\over Dt}
+ \tilde{\Sigma}~ {\bf \nabla}\cdot
\tilde{\bf u} = 0~,
\eqno(1a)
$$
$$
{D \tilde{\bf u}\over Dt}  =
-{1\over \tilde{\Sigma}}
{\bf \nabla}\tilde{P}
 - {\bf \nabla}\Phi ~,
\eqno(1b)
$$
$$ 
{D S\over Dt} = 0~,
\eqno(1c)
$$
where $D/Dt \equiv \partial /\partial t
+ \tilde{\bf u}\cdot {\bf \nabla}$, and where
 $S \equiv {\tilde P}/(\tilde \Sigma)^\gamma$
is the entropy of the disc matter.

We consider perturbations $\sim f(r){\rm exp}
(im\phi - i \omega t)$, where $m=0,1,2,..$ is the azimuthal
mode number and $\omega$ the angular frequency.
For free perturbations $\omega=\omega_r+i\omega_i$ and for
the growing modes of interest $\omega_i >0$. 
         From equation (1a), we have
$$i\Delta \omega~ \delta \Sigma =
{\bf \nabla} \cdot
(\Sigma ~\delta {\bf u})~,
\eqno(2)
$$
where
$$
\Delta \omega(r) \equiv \omega - m \Omega(r)~,
$$
and $\Omega = u_\phi/r$.

 From equation (1b) we have
$$ 
i\Delta \omega \delta u_r +
2 \Omega \delta u_\phi =
{1\over \Sigma}{\partial \delta P \over \partial r}
-{\delta \Sigma \over \Sigma^2} {d P\over dr}+\nabla\delta \Phi~,
\eqno(3a)
$$
$$
i\Delta \omega \delta u_\phi -
{\Omega_r^2 \over 2 \Omega}
\delta u_r = ik_\phi
 {\delta P\over \Sigma}+ik_\phi \delta \Phi
\eqno(3b)
$$
Here,
$\Omega_r \equiv
[r^{-3} d(r^4 \Omega^2)/dr]^{1\over 2}$
 is the radial
epicyclic frequency, and $k_\phi \equiv m/r$
is the azimuthal wavenumber.  For an approximately
Keplerian disc, $\Omega_r \approx \Omega$.

    From equation (1c) and (1d), we have
$$
\delta P =c_s^2 \delta \Sigma 
-{i\Sigma c_s^2 \over \Delta \omega L_S}\delta u_r~.
$$
Here, $c_s = (dP/d\Sigma)_S^{1/2}$ is
the effective sound speed in the disc and
 $L_S^{-1} \equiv \gamma^{-1}d\ln(S)/dr $ with
$L_S$ the length-scale of the entropy 
$S=P/\Sigma^\gamma$  variation in the disc.  
To simplify the subsequent calculations we consider
the homentropic case where $L_S\rightarrow \infty$.

   The perturbation of the gravitational potential is
give by
$$
\nabla^2 \delta \Phi = 4\pi G \delta \Sigma \delta(z) ~.
\eqno(4)
$$

   Equations (3) can be solved to give
$$
     \Sigma \delta u_r = iF\left[ {\Delta \omega \over \Omega}
     {\partial \delta \Psi \over \partial r} -2 k_\phi \delta \Psi\right]~,
 \eqno(5a)
 $$
 $$ 
~~~~\Sigma \delta u_\phi =F\left[ {\Omega_r^2 \over 2 \Omega^2}
 {\partial \delta \Psi \over \partial r} -k_\phi {\Delta \omega \over \Omega}
 \delta \Psi \right]~,
 \eqno(5b)
 $$
where 
$$\delta \Psi \equiv c_s^2{\delta \Sigma\over \Sigma} + \delta \Phi~,
\eqno(5c)
$$
 is an effective
potential and   
$$
F \equiv {\Sigma \Omega\over \Omega_r^2 -(\Delta \omega)^2}~,
\eqno(5d)
$$
is the function identified by Lovelace and Hohlfeld (1978).

   Substituting equations (5) into (2) we obtain
$$
\delta \Sigma ={1\over r}{\partial \over \partial r}
\left({rF\over \Omega}{\partial \delta \Psi\over \partial r}\right)
-k_\phi^2 {F \over \Omega} \delta \Psi
- {2k_\phi \over \Delta \omega} {dF\over dr} \delta \Psi~,
\eqno(6) 
$$

 \subsection{Radially Localized Modes}
 
     Here we consider radially localized modes in
the sense that perturbation extends over a radial
region $\Delta r$ with $(\Delta r)^2 \ll r^2$ centred
at $r_0$.
With this in mind we rewrite equation (6) as
$$ (\Omega_r^2 - \Delta \omega^2) {\delta \Sigma\over \Sigma}
={d^2\delta \Psi \over dr^2}- k_\phi^2 \delta \Psi~~~~~~ ~~~~~~~~~~~~~~~~~~~~~~
$$
$$
~~~~~~~~~~~~~~~~~~~~~~~+{d\ln \tilde F \over dr}
{d\delta \Psi \over dr} -{2k_\phi \Omega \over \Delta \omega}
{d \ln F \over dr} \delta \Psi~,
\eqno(7)
$$
where $\tilde F \equiv rF/\Omega$.

  We let 
$$
(\delta \Sigma_{k_r},~ \delta \Psi_{k_r})=\int {dr\over 2\pi}(\delta \Sigma,~\delta \Psi)
\exp[-ik_r(r-r_0)]~,
$$
where $k_r$ is the radial wavenumber of the perturbation.
The radial Fourier transform of equation (7)
gives 
$$
(\Omega_r^2-\Delta \omega^2) {\delta \Sigma_{k_r} \over \Sigma} 
= -{\bf k}^2 \delta \Psi_k  ~~~~~~~~~~~~~~~~~~~~~~~~~~~~~~~
$$
$$
~~~~~~~~~+
\left({d\ln \tilde F \over dr}
{d\delta \Psi \over dr} -{2k_\phi \Omega \over \Delta \omega}
{d \ln F \over dr} \delta \Psi\right)_{k_r}~.
\eqno(8)
   $$
 Here, we have neglected the radial variation of
 $\Sigma$ and $\Omega_r$, and that of  $\Delta \omega^2$
 in comparison with $\Omega_r^2$.  Also, ${\bf k}=
 k_r\hat{\bf r} +k_\phi \hat{\rvecphi~}$ and $(..)_{k_r}$ 
 denotes the Fourier transform.
 
      Assuming $|k_r| \sim
 (\Delta r)^{-1} \gg r_0$,  the WKBJ solution of equation (4)   gives
$\delta \Phi_{k_r} = - 2\pi G \delta \Sigma_{k_r}/|{\bf k}|$.
Therefore, from equation (5c) we obtain
$$
\delta \Psi_{k_r} =\left(1-{2k_c \over |{\bf k}|}\right) c_s^2 {\delta \Sigma_{k_r} \over \Sigma}~,
\eqno(9a)
$$
where 
$$
k_c \equiv {\pi G \Sigma \over c_s^2}~
\eqno(9b)
$$
is a characteristic wavenumber.

\begin{figure}
\centering
\includegraphics[scale=0.4]{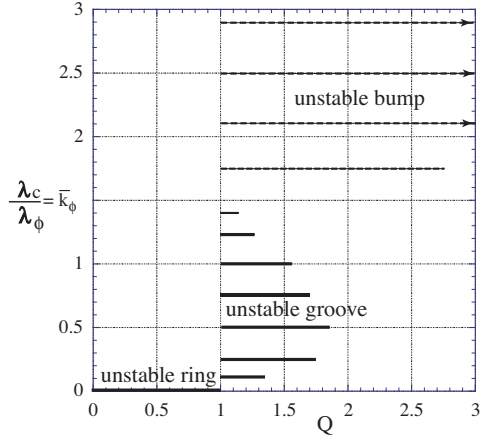}
\caption{The horizontal lines mark out the
regions of `bump' and `groove' instability
in a self-gravitating disc for $m=3$ and the
ring instability for $m=0$.  Here,
$\lambda_c=2\pi/k_c$ and $\lambda_\phi =2\pi/k_\phi$.
For the $Q>1$ part of the figure we have
used equations (12) and (15) with 
$\Delta F=0.2$ for the bump and $\Delta F=-0.2$ for
the groove, $\Delta x =0.04$, and $q=3/2$.}
\end{figure}
   
\subsubsection{Axisymmetric Modes}

       For axisymmetric perturbations ($k_\phi = 0 = m$) of
smooth disc (where the term $\propto ~ d \ln\tilde F/dr$ 
can be neglected in equation (8) ), equations (8) and (9a)
give
$$
\omega^2 = \Omega_r^2+k_r^2c_s^2 -2k_c|k_r|~,
\eqno(10)
$$
(Safranov 1960; Toomre 1964).
The minimum of $[\omega(k_r)]^2$ occurs at $k_r =k_c$
where $\omega^2= \Omega_r^2 -k_c^2 c_s^2$.   
   Therefore, if we define $k_{c*} \equiv \Omega_r/c_s$,
then for $k_c <k_{c*}$ the minimum of $\omega^2$ 
is positive and the axisymmetric perturbations are stable.
Conversely for $k_c > k_{c*}$ the perturbations are
unstable.     With
$$
 Q \equiv {k_{c*} \over k_c} = {\Omega_r c_s \over \pi G \Sigma}~,
 \eqno(11)
 $$
 the axisymmetric perturbations are stable (unstable) for
 $Q>1$ ($Q<1$) (Toomre 1964).    
    The minimum value
 of the squared frequency is  
 $\omega(k_r)^2 =(k_{c*} c_s)^2(1-Q^{-2})$.
 Figure 1 shows the unstable region for a ring,
 $Q<1$.
 
 \subsubsection{Non-Axisymmetric Modes}
 
    For the non-axisymmetric perturbations ($m=1,2,..$, $k_\phi=m/r$)
we assume that $(\Delta \omega)^2 \ll \Omega_r^2$ in equation (8).
   Multiplying equation (8) by $(1-2k_c/|{\bf k}|)$ and doing the inverse
Fourier transform gives
$$
\delta \Psi(x) ={2\over \pi} \int_{-\infty}^{\infty} dx^\prime
K(x-x^\prime){\Omega \over \Delta \omega(x^\prime)}
{d \ln F \over dx^\prime} \delta \Psi(x^\prime)~.
\eqno(12a)
$$
 This represents an integral equation for $\delta \Psi(x)$.  
     The term in
 equation (8) proportional to $d\ln \tilde F/dr$ has been
 dropped.  It is negligible compared with
 the term involving $ d \ln F/dr$ because $|m\Omega/\Delta \omega |
 \gg 1$ for the considered modes.
      The kernel is
$$
K(x) = {2\over \pi}\int_0^\infty dy {\cos(m x y)
\left({2\over \overline{k}_\phi \sqrt{1+y^2}} -1 \right)
\over
\left(1+y^2+{Q^2 \over \overline{k}_\phi^2}- 
{2 \sqrt{1+y^2}\over \overline{k}_\phi}\right)}~.
\eqno(12b)
$$
Here, $x \equiv (r-r_0)/r_0$ with $x^2 \ll 1$, $\Omega=
\Omega(r_0)$, 
$Q$ is given in equation (11), and
$\overline{k}_\phi \equiv k_\phi/k_c$.

   The half-thickness of the disc $h$
is  $c_s/\Omega_r$ so that
$$
{\overline{k}_\phi\over m Q} ={h\over r}~,
\eqno(13)
$$
where we assume $(h/r)^2 \ll 1$.   
  Note that $\lambda_c = 2\pi/k_c = 2\pi h Q$.
The Toomre (1964) characteristic wavelength
is defined as $\lambda_*=4\pi G\Sigma/\Omega_r^2 =4 h/Q$.
Thus $\lambda_c/\lambda_* =\pi Q^2/2$.

     In parallel with equations (12) note that
  $$
c_s^2{\delta \Sigma(x)
\over \Sigma} ={2\over \pi} \int_{-\infty}^{\infty} dx^\prime
G(x-x^\prime){\Omega \over \Delta \omega(x^\prime)}{d \ln F \over dx^\prime} \delta \Psi(x^\prime)~.
\eqno(14a)
$$
    Here,
$$
G(x) = -{2\over \pi}\int_0^\infty dy {\cos(m x y)
\over
\left(1+y^2+{Q^2 \over \overline{k}_\phi^2}- 
{2 \sqrt{1+y^2}\over \overline{k}_\phi}\right)}~,
\eqno(14b)
$$
is a second kernel.

     We solve equation (12) for the case where
 $\ln [F(x)]$ has a square bump or groove.  That is,
 $$
\ln [F(x) ]= \Delta F H(x+\Delta x/2) H(\Delta x/2 -x)~,
\eqno(15)
 $$
  where $H$ is the Heaviside step function ($+1$ for positive
  argument and zero for a negative argument), and $\Delta x$
  is the fractional radial width of the bump ($\Delta F >0$)
  or groove ($\Delta F<0$).
  This distribution of $\ln F$ is known to be useful for
analytic calculations of the non-axisymmetric instability
of  discs (Sellwood \& Kahn 1991; Umurhan 2010).

Substituting equation (15) into (12) gives
$$
\delta \Psi_-={2\over \pi}K_0{\Omega \Delta F\over \Delta \omega_-}
\delta \Psi_- 
-{2\over \pi} K_{\Delta x}{\Omega \Delta F\over \Delta \omega_+}\delta \Psi_+~,
\eqno(16a)
$$
$$
\delta \Psi_+={2\over \pi} K_{\Delta x}{\Omega \Delta F\over \Delta \omega_-}\delta \Psi_- -{2\over \pi}K_{0}{\Omega \Delta F\over \Delta \omega_+}
\delta \Psi_+~,
\eqno(16b)
$$
where $K_0 =K(0)$, $ K_{\Delta x}=K(\Delta x)$
and $\Delta \omega_\pm =\omega -m\Omega[r_0(1 \pm \Delta x/2)]
\approx \omega-m\Omega \pm q m \Delta x\Omega/2$ for
the $q \equiv - d\ln\Omega/d\ln r$ with $q=3/2$ for a Keplerian
disc.
   Equations (16) give a quadratic in $\omega$ which for
the assumed symmetry of the bump or groove about $x=0$ gives
either  $(\omega-m\Omega)^2  >0$ - stable motion - or
$(\omega-m\Omega)^2  <0$  - unstable motion.

     Evidently an arbitrary $\ln[ F(r)]$ can be approximated by
a staircase-like function with the result that equation (12)
leads to an $n\times n$ matrix equation in place of equation (16)
with $n$ the number of steps.
   
\section{Sample Results}
   
   The main  parameters of importance for instability are:  
   (1) The parameters of the bump or groove ($\Delta F$ and $\Delta x$).  (2) The parameters of the disc $Q$ and of the perturbation $m$ and $\overline{k}_\phi =k_\phi/k_c$.    
   Figure 1 gives a qualitative picture of the regions of bump, groove, and ring instability in a disc as a function of $Q$ and $\overline{k}_\phi = k_\phi/k_c$.  The needed kernel  in equation (12b) is accurately
 evaluated numerically using a $4000$ point integration.
       (Our attempt to analytically integrate (12b) was thwarted by
the required branch cuts at the points $y=\pm i$.)
   The transition from groove to bump instability occurs when $K(0)$
goes through zero as a function of $\overline{k}_\phi $, with
$K(0)>0$  for small $\overline{k}_\phi $. 
    We find that this occurs at $\overline{k}_\phi \approx 1.55$.
    This agrees with the conclusion of Lovelace and
Hohlfeld (1978)  that the transition is at
$\overline{k}_\phi =\pi/2 \approx 1.57$.

\begin{figure}
\centering
\includegraphics[scale=0.4]{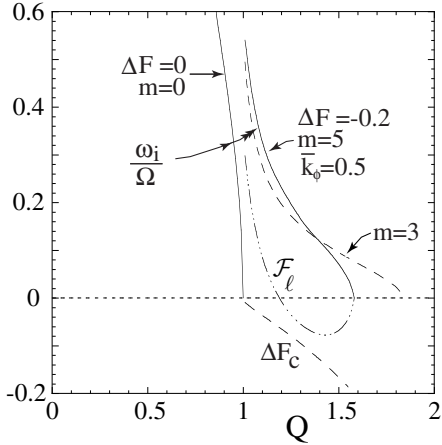}
\caption{Growth rates of axisymmetric perturbations ($m=0$)
for $Q<1$ and non-axisymmetric perturbations ($m=3,~5$) for
$Q>1$ for a groove ($\Delta F <0$) in the $F(r)$ profile,
$\overline{k}_\phi = \lambda_c/\lambda_\phi =0.5$, and
$q=3/2$.
Here, $\Delta F_c$ is the threshold value of the depth
of the groove $\Delta F$ needed for instability.  Also, ${\cal F}_\ell$
is proportional to the radial angular momentum flux across the centre
of the groove as discussed in the text.
For the $Q>1$ part of this figure we have
used equations (12) and (15) with 
$\Delta F=-0.2$ for
the groove and $\Delta x =0.04$.}
\end{figure}

    Figure 2 shows the growth rates as a function of $Q$ for
the ring mode ($m=0$ and $Q<1$) and the groove mode
($m=3,~5$, $Q>1$, and $\overline{k}_\phi =0.5$).   Also 
shown in the figure is $\Delta F_c$ which is the minimum
groove depth needed for instability.    Further,
${\cal F}_\ell$ is proportional to the radial angular momentum flux
across the $x=0$ surface. 
    Specifically, ${\cal F}_\ell \propto r \int d\phi
\Re(r^2\Omega \delta \Sigma^*\delta u_r + \Sigma r \delta u_\phi^*
\delta u_r)$, where $\Re$ denotes the real part.

   Generally, for both unstable grooves and bumps the growth
rate increases as $|\Delta F|$ increases {\it above} a threshold value.
Also, the growth rate decreases as the width of the groove
or bump $\Delta x$ increases.  
 This supports the interpretation of Umurhan (2010) that the
instability of a bump arises from interaction ``edge waves''
on the surfaces where $F$ changes rapidly.

  {\bf Unstable grooves:}  Figure 3 shows the   density perturbation  $\delta \Sigma(r,\phi)$ and the arrows represent the flow perturbation  $(\delta u_r,~\delta u_\phi)$
 for an unstable groove  with $\Delta F=-0.2$, and $Q=1.3$, $m=3$, and $\overline{k}_\phi =0.5$.   Notice that the anticyclonic motion of
 the vortex about the {\it minimum} of the density perturbation.

\begin{figure}
\centering
\includegraphics[scale=0.4]{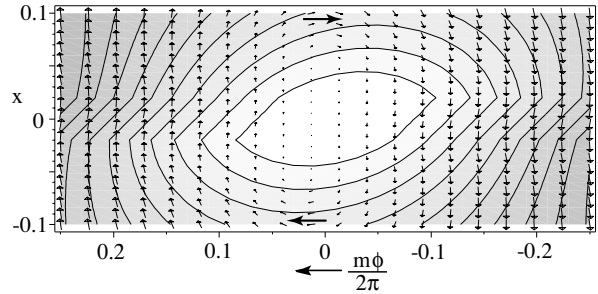}
\caption{The  background gray-scale (white=low, dark=high) 
and the contours represent
the density perturbation $\delta \Sigma(r,\phi)$, and the arrows
represent the flow perturbation  $(\delta u_r,~\delta u_\phi)$
for the case of an unstable groove,  $\Delta F=-0.2$, and $Q=1.3$, $m=3$,  $\overline{k}_\phi =0.5$, and $q=3/2$.
 The growth rate is $\omega_i/\Omega =0.152$.  The larger arrows near the top and bottom of the figure have
been put in to make clear the direction of the circulation which is
anticyclonic.  In this case this circulation is around a minimum of
the density perturbation.}
\end{figure}

\begin{figure}
\centering
\includegraphics[scale=0.4]{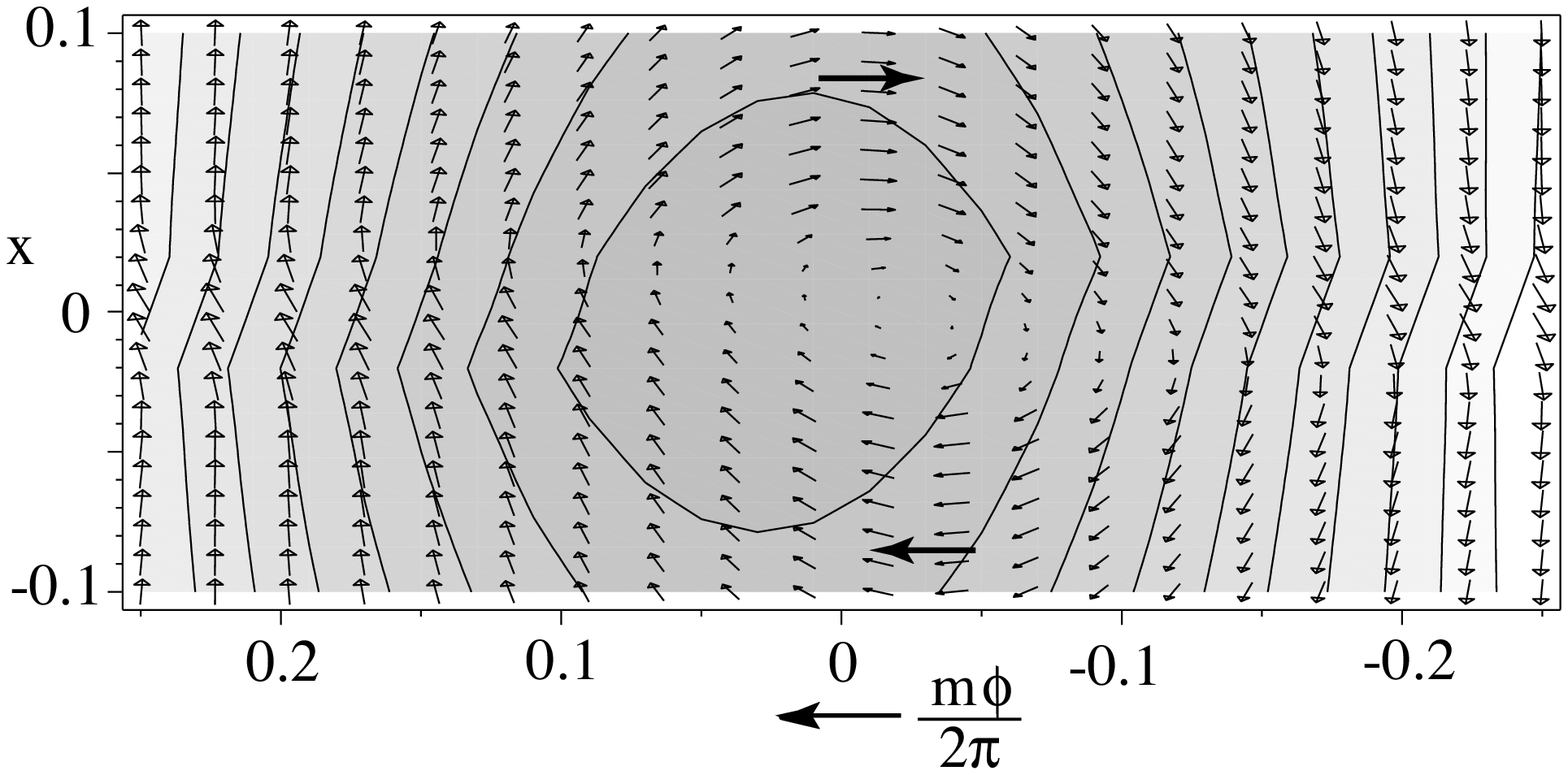}
\caption{The  background gray-scale (white=low, dark=high) and the contours represent
the density perturbation $\delta \Sigma(r,\phi)$, and the arrows
represent the flow perturbation  $(\delta u_r,~\delta u_\phi)$
for the case of an unstable bump, $\Delta F=0.2$, and
 $Q=1.3$, $m=3$,  $\overline{k}_\phi =2$, and $q=3/2$.
   The growth rate is $\omega_i/\Omega =0.0844$.
   The larger arrows near the top and bottom of the figure have
been put in to make clear the direction of the circulation which is
anticyclonic.  In this case this circulation is approximately
around a maximum of
the density perturbation.}
\end{figure}

\begin{figure}
\centering
\includegraphics[scale=0.4]{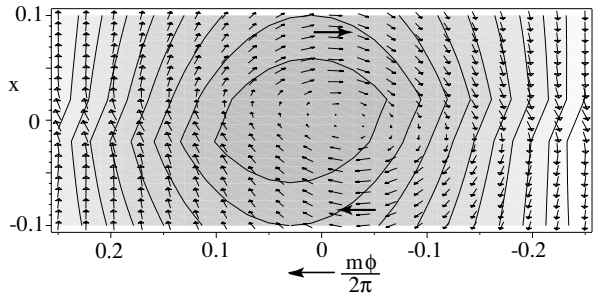}
\caption{The  background gray-scale (white=low, dark=high) and the contours represent
the density perturbation $\delta \Sigma(r,\phi)$, and the arrows
represent the flow perturbation  $(\delta u_r,~\delta u_\phi)$
for the case of an unstable bump, $\Delta F=0.2$, and
 $Q=2$, $m=3$,  $\overline{k}_\phi =2$, and $q=3/2$.
   The growth rate is $\omega_i/\Omega =0.0661$.
      The larger arrows near the top and bottom of the figure have
been put in to make clear the direction of the circulation which is
anticyclonic.  In this case this circulation is approximately
around a maximum of
the density perturbation.}
\end{figure}

    {\bf Unstable bumps:}   Figure 4  shows the   density perturbation  $\delta \Sigma(r,\phi)$ and the arrows represent the flow perturbation  $(\delta u_r,~\delta u_\phi)$
 for an unstable bump  with $\Delta F=0.2$, and $Q=1.3$, $m=3$, and $\overline{k}_\phi =2$.   Notice that the center of the anticyclonic motion of the vortex is  offset  by a small angle from the {\it maximum}  of the density perturbation.    
        The angular momentum flux across $x=0$ is found
 to be always outward for unstable bumps.  
 
    Figure 5  shows a further case of an unstable bump with 
$\Delta F=0.2$, and $Q=2$, $m=3$, and $\overline{k}_\phi =2$. 
     The center of the anticyclonic motion of the vortex is again offset  by a small angle from the {\it maximum}  of the density perturbation.

 \section{NUMERICAL EXAMPLES} 
   
        We first consider an accretion disc around a solar mass star with
 accretion rate  $\dot{M}=10^{-8} M_\odot$yr$^{-1}$.      
  For an
 $\alpha-$viscosity disc with $\alpha =5\times10^{-3}$, the surface mass
 density is $\Sigma =89.3$~gcm$^{-2}$ at $r=10$AU assuming the half-thickness of the disc is $h=0.04 r $ or $c_s/v_K =0.04$, where
 $c_s$ is the isothermal sound speed in the disc  and $v_K = 9.4$ kms$^{-1}$ is the Keplerian velocity.   We have used $\dot{M}=2\pi r \Sigma
 u_r$ with $u_r =\alpha c_s^2/v_K$.
     For comparison, the surface mass-density of a Jupiter mass spread over a radial extent $2h$ at $10$AU is
about $170$ gcm$^{-2}$.
     From equations (9b)  and (10), $k_c = 1.97/r$ and $Q=12.7$.     
For fixed $c_s/v_K$, notice that $Q \propto \dot{M}/\alpha$.
     In equation (13)
we have $\overline{k}_\phi=0.04mQ=0.508m$.   
    For $\overline{k}_\phi < \pi/2$
 or $m \leq 3$ there is no instability because $Q\gg 1$.
     We have a bump instability for $m>3$.   
Clearly, all of the modes $m\ge 1$ are in the bump regime if $Q>\pi r/(2h)$.    
    The mass of the disc is relatively small,
 $\sim \pi r^2 \Sigma/M_\odot =(c_s/v_K)/Q=0.0031$, so that it is approximately Keplerian.
 
         As a second case, consider a  larger
 accretion rate  $\dot{M}=3\times 10^{-8} M_\odot$yr$^{-1}$
 and a smaller viscosity $\alpha=0.002$
 but with the other quantities ($r$,  $c_s$) the same as above.
 Then we find that $Q=1.69$ and $\overline{k}_\phi=0.0676m$.   
       For $\overline{k}_\phi < \pi/2$
 or $m \leq 23$ there can  be a groove instability because $Q$ is not much
 larger than unity (see Fig. 1).  At the same time we expect  the bump instability
for $m>23$  to be unimportant because the azimuthal wavelengths
are comparable to the disc half-thickness.

 \section{Conclusions}

     We have analyzed the Rossby wave instability 
in a continuum of thin disc models ranging 
from self-gravitating to non-selfgravitating.  
    The important quantities
determining the stability/instability are: (1) the parameters of the bump
(or depression) in the inverse potential vorticity $F(r)$ at $r_0$, (2) the Toomre
$Q$ parameter of the disc, and (3) the dimensionless azimuthal wavenumber of the  perturbation $\overline{k}_\phi =mQh/r_0$, where $h$ is the 
half-thickness of the disc and $m=1,~2,..$ is the azimuthal mode number.
     For  $\overline{k}_\phi  <  \pi/2$ {\it and} $Q\lesssim 2$
instability may occur for  the case of
a  groove in the surface-density.  In this case
the centres of the anticyclonic vortices have reduced density.     
    For  $\overline{k}_\phi  <  \pi/2$ {\it $Q>2$}, both bumps
and grooves are stable.
  For  $\overline{k}_\phi  >  \pi/2$,  the bumps may be unstable
and the centres of the anticyclonic vortices have density enhancements.
     The growth rates for both the groove and bump instability are 
$\lesssim 10\%$ of the orbital angular frequency for bumps fractional
amplitude $\lesssim 20\%$.   

    For fixed radius $r_0$ and sound speed $c_s$,  the regime
of instability - groove or bump - is set by the $Q$ value which
is proportional to $\dot{M}/\alpha$ with $\alpha$ the 
dimensionless viscosity.   For $Q\lesssim 2$ and  
$m < (\pi/2)(r_0/h)Q^{-1}$ the groove 
instability may  occur.   
   On the other hand  for $Q\gg 1$ 
and $m>(\pi/2)(r_0/h)Q^{-1}$ the bump instability may occur.
      The estimates of Sec. 4 indicate that the two regimes are
 not widely separated in terms of $\dot{M}$ and $\alpha$.  
 For this reason we suggest that simulations of  Rossby vortices
 include the self-gravity for cases where $Q< r_0/h$.

\section*{Acknowledgements}

  We thank the referee for helpful criticism.
RVEL was supported in part by NASA grant NNX11AF33G.


\begin{thebibliography}{0}


\bibitem{} Barge, P., \& Sommeria, J. 1995, A\&A, 295, L1

\bibitem{} Bracco, A., Chavanis, P.H., Provenzale, A., \& Spiegel,
E.A. 1999, Phys. Fluids, 11, 2280

\bibitem{} Crespe E., Gonzalez J.-F., \&  Arena S. E., 2011, in SF2A-2011:
Proceedings of the Annual meeting of the French Society
of Astronomy and Astrophysics, G. Alecian, K. Belkacem,
R. Samadi, \& D. Valls-Gabaud, ed., pp. 469Ð473


\bibitem{} Gammie, C.F. 1996, ApJ, 457, 355

\bibitem{} Gammie, C.F. 2001, ApJ, 553, 174

\bibitem{} Godon, P., \& Livio, M. 2000, ApJ, 537, 396


\bibitem{} Heng, K., \& Kenyon, S.J. 2010, MNRAS, 408, 1476

\bibitem{} Johanssen, A., Andersen, A.C., \& Brandenburg, A.
2004, A\&A, 417, 361

\bibitem{} Li, H., Finn, J.M., Lovelace, R.V.E., \& 
Colgate, S.A. 2000, ApJ, 533, 1023

\bibitem{} Li, H., Colgate, S.A., Wendroff, B., \& Liska, R.
2001, ApJ, 551, 874

\bibitem{} Lin, M.-K., 2012, ApJ, 754, 21





\bibitem{} Lovelace, R.V.E., \& Hohlfeld, R.G. 1978, ApJ, 221, 51



 \bibitem{} Lovelace, R.V.E., Li, H., Colgate, S.A., \&
Nelson, A.F. 1999, ApJ, 513, 805

\bibitem{} Lovelace R. V. E., Turner L., Romanova M. M., 2009, ApJ,
701, 225



\bibitem{} Lyra, W., Johansen, A., Zsom, A., Klahr, H., \& Piskunov, N.
2009, A\&A, 497, 869



\bibitem{} Meheut, H., Casse, F., Varni\`ere, P., \& Tagger, M. 2010,
A\&A, 516, A31

\bibitem{} Meheut, H., Keppens, R., Cassee, F., \& Benz, W. 2012a,
A\&A, 542, A9

\bibitem{} Meheut, H., Yu, C., \& Lai, D. 2012b,  MNRAS, 422, 2399

\bibitem{} Meheut, H., Lovelace, R.V.E., \& Lai, D. 2012c,
MNRAS, in press

\bibitem{} Papaloizou, J. C. B., \& Pringle, J. E. 1984, MNRAS, 208, 721;
------- 1985, MNRAS, 213, 799

\bibitem{} Safronov, V.S. 1960, Ann. Astrophys., 23, 982

\bibitem{} Sellwood, J.A. 2012, ApJ, 751, 44

\bibitem{} Sellwood, J.A., \& Kahn, F.D. 1991, MNRAS, 250, 278

\bibitem{} Varni\`ere, P., \& Tagger, M. 2006, A\&A, 446, L13


\bibitem{} Tagger, M., \& Pellat, R. 1999,  A\&A, 349, 1003

\bibitem{} Tagger, M., \& Varni\`ere, P. 2006,  ApJ, 652, 1457

\bibitem{} Tagger, M., \& Melia, F. 2006, ApJL, 636, L33

\bibitem{} Tanga, P., Babiano, A., Dubrulle, B., \& Provenzale, A. 1996,
Icarus, 121, 158

\bibitem{} Toomre, A. 1964, ApJ, 139, 1217

\bibitem{} Umurhan, O.M. 2010, A\&A, 521, A25




\end{thebibliography}
\end{document}